\documentclass[manuscript,screen]{acmart}

\AtBeginDocument{%
  }

\begin{document}

\title{Dynamic Question-Answering of Clinical Documents using Retrieval Augmented Generation}

\author{Ran Elgedawy}
\email{relgedaw@vols.utk.edu}
\affiliation{%
  \institution{University of Tennessee, Knoxville}
   \country{USA}
}
\author{Ioana Danciu}
\email{danciui@ornl.gov}
\affiliation{%
  \institution{Oak Ridge National Laboratory}
   \country{USA}
}
\author{Maria Mahbub}
\email{mahbubm@ornl.gov}
\affiliation{%
  \institution{Oak Ridge National Laboratory}
   \country{USA}
}
\author{Sudarshan Srinivasan}
\email{srinivasans@ornl.gov}
\affiliation{%
  \institution{Oak Ridge National Laboratory}
   \country{USA}
}

\renewcommand{\shortauthors}{Elgedawy et al.}

\begin{abstract}
Electronic health records (EHRs) contain a wealth of valuable patient information embedded within clinical notes. The ever-increasing volume and complexity of these unstructured clinical notes however, make it extremely challenging for researchers and clinicians to manually locate and extract relevant information in a timely manner. Intelligent systems that can understand clinical documents and enable users to efficiently query and retrieve key details on demand are urgently needed. In this work we present a natural language conversational interface powered by state-of-the-art open source large language models (LLMs) that allows users to explore clinical notes through dynamic question-answering. We built a chatbot interface using Langchain, a framework for creating and managing applications powered by large language models, paired with powerful transformer-based language models. The system allows users to simply ask questions in natural language and receive answers extracted from relevant parts of the clinical notes. We evaluated combinations of semantic embedding models (EM) such as the state of the art SentenceTransformers framework, based on the  Sentence-BERT EM\cite{reimers-2019-sentence-bert}, and advanced large language models on their ability to encode both queries and documents for optimal information retrieval. Wizard Vicuna demonstrated the highest accuracy due to its 13B parameters but required heavy compute capabilities. To improve inference latency and deployability, we used model optimization techniques like weight quantization to reduce latency by approximately \emph{48} times. While results are promising, several limitations remain including model hallucinations, and lack of robust evaluation across diverse medical cases. Addressing these gaps represents important future work to unlock the full potential of clinical notes and advance clinical decision making through artificial intelligence (AI).\end{abstract}




\maketitle

\section{Introduction}
Electronic health record (EHR) systems have become ubiquitous in modern healthcare environments \cite{ehr1, article, manuel2010importance}. They offer an integrated digital repository of patient health information collected from various clinical encounters across time. EHRs may contain  demographics, medical and family histories, laboratory test results, radiology images, computed biosignals and crucially, clinical notes generated by physicians, nurses and other care team members. These clinical notes, documenting interactions with the healthcare system, encapsulate a wealth of vital information about patients' conditions, treatments, care plans and outcomes in rich natural language. However, a significant limitation of EHRs is that critical patient information is buried within large volumes of unstructured clinical notes that are challenging for providers to efficiently search through and extract relevant details on demand \cite{10.1371/journal.pdig.0000347,Sedlakova2022.07.28.22278137}. For instance, a newly assigned physician may need to rapidly comprehend a patient's care history across multiple specialists to make appropriate decisions \cite{4b050ffe-544d-33dd-8cf9-852eb6c8b9b8}, or a researcher may want to analyze treatment response patterns across populations \cite{doi:10.7326/M20-0104}. These use cases require intelligently sifting through troves of notes to pinpoint key facts, relationships and temporal trends - a formidable manual task. Hence, there is a profound need for intelligent systems that can automatically understand clinical notes written in natural language and enable users to easily query and retrieve targeted information on the fly through conversational interactions.

Traditionally, information retrieval systems have relied on methods like regular expression matching \cite{Ilie2016}, Boolean logic \cite{Melucci2009}, probabilistic models \cite{10.1145/582415.582416}, and term frequency-inverse document frequency (TF-IDF) \cite{ref1} for surface-level matching of keywords between queries and documents. While these techniques have been useful for basic keyword-based retrieval, they have their limitations. Boolean models \cite{Melucci2009} utilize operators like AND, OR, and NOT to retrieve documents containing specific keyword combinations, but they lack the ability to understand the nuances and context of natural language queries. Probabilistic models \cite{10.1145/582415.582416} rank results based on keyword occurrences, which can be effective for certain tasks, such as information retrieval in search engines \cite{sontag2012probabilistic}, spam email detection \cite{lau2012text}, and sentiment analysis in text data \cite{celikyilmaz2010probabilistic}, but they struggle to grasp the underlying meaning of the text. TF-IDF \cite{ref1} is another widely used method that assigns weights to keywords in queries and documents based on their rarity. This approach helps identify important terms, but it still falls short in capturing the full semantic context and relationships between words. The shared limitations across these approaches are the lack of semantic understanding and the ability to interpret the context holistically.

As a result, these traditional techniques often miss relevant information when applied to complex natural language text. They fail to recognize synonyms, antonyms, word variations, context, and other language intricacies that are essential for accurate and comprehensive information retrieval.

Large language models (LLMs) based on the transformer architecture \cite{NIPS2017_3f5ee243} can be used to overcome these limitations and unlock the deeper meaning. The rapidly advancing LLMs, like GPT-4 \cite{openai2023gpt4}, have revolutionized the field of natural language processing, with their contextual understanding, generative capabilities, and fine-tuning adaptability \cite{zhao2023survey,Hadi_2023}. Integrating LLMs into information retrieval systems holds the potential to deliver more accurate, insightful, and context-aware results, transforming the way users access information in a more personalized and efficient manner.  

Embedding models (EMs) serve as powerful tools in natural language processing by representing words or phrases in a continuous vector space where the geometric distances between vectors capture semantic relationships between words. EMs, such as word embeddings \cite{almeida2023word}, play a crucial role in capturing intricate relationships and contextual nuances within the text.

Retrieval augmented generation (RAG) refers to a novel technique where a model first retrieves relevant information from a large corpus or dataset and then generates responses or outputs based on the retrieved information. This approach combines the benefits of both retrieval-based and generation-based methods, allowing for more contextually relevant and informative responses in conversational systems. Recent research works are exploring the trade-offs between this method and traditional fine-tuning approaches for language models, with some favoring RAG for certain contexts \cite{greyling2021rag, balaguer2024rag}.

In this work, we explore the utility of modern natural language processing (NLP) methods based on a combination of Large Language Models (LLMs), Retrieval Augmented Generation (RAG), and embedding models to develop a conversational agent that enables users to interactively explore clinical notes through dynamic question answering. By 'dynamic question answering,' we mean that users can engage with the system in a flexible manner, posing inquiries about clinical notes in natural language. By interfacing with the agent akin to interacting with a human assistant, users can extract pertinent details buried within notes, thereby facilitating improved clinical care and research endeavors.

Our work serves as a demonstration of practical applicability, particularly in information extraction from clinical notes and the development of chatbots utilizing this documentation. Highlighting its relevance in assisting users or researchers in swiftly and effectively extracting pertinent information from clinical notes, our work demonstrates the practical applicability of LLMs in the medical domain, particularly in information extraction from clinical notes and the development of chatbots utilizing this documentation. We highlight the significance of LLMs in assisting users to swiftly and effectively extract pertinent information from clinical notes. 

The following are our contributions:
\begin{itemize}\itemsep=1mm
\item 
We developed a conversational question answering system based on LLMs that allows natural language interaction for exploring clinical notes
\item
We conducted experiments evaluating combinations of different EM and LLMs to enhance performance and speed
\item 
We demonstrated the advantages of weight quantization and the superiority of retrieval augmented generation over domain-specific fine tuning
\end{itemize}

\textbf{Organization: }The paper is organized as follows: We discuss related works in section \ref{sec: rw}. Section \ref{sec:approach} describes the approach. Details of the dataset are available in section \ref{sec:data}. Evaluation and results are presented in section \ref{sec:results}. We discuss the limitations and future work in section \ref{sec:lf} and section \ref{sec:conc} concludes the paper.

\section{Related Work} \label{sec: rw}
\subsection{LLMs in information retrieval}
The intersection of LLMs and Information Retrieval (IR) has emerged as a pivotal area of research, transforming traditional approaches and reshaping the landscape of information access. 


The foundational work of \citeauthor{zhu2023large} \cite{zhu2023large} offers a comprehensive survey covering diverse applications of LLMs in information retrieval (IR). This work spans various facets of IR, including query rewriting, retrieval, re-ranking, and reading. Additionally, an influential contribution in the information retrieval domain is the tutorial by \citeauthor{10.1145/3477495.3532681} \cite{10.1145/3477495.3532681} . This tutorial advances both theoretical and practical aspects of transformer-based approaches, emphasizing efficiency and robustness in large-scale information retrieval scenarios. Its goal is to provide the audience with insights gained from developing and fine-tuning transformer/BERT-based models for IR tasks.

In contrast, our approach focuses on the utilization of off-the-shelf models, enabling the direct retrieval of information from provided resources without the need for intricate fine-tuning procedures. This streamlined method significantly boosts usability and efficiency, circumventing the time and computational expenses associated with fine-tuning. Furthermore, our research demonstrates that in our case the performance achieved through our direct utilization method surpasses that attained through domain-specific fine-tuning processes. 

\subsection{Medical question answering using LLMs}

In recent years, the field of medical question answering has experienced remarkable advancements, fueled by the integration of LLMs. Various research works have significantly contributed to the pursuit of expert-level performance in this challenging domain. \citeauthor{10.1093/bioinformatics/btz682} \cite{10.1093/bioinformatics/btz682} addresses the unique challenges of biomedical text mining by introducing BioBERT, a domain-specific language representation model pre-trained on large-scale biomedical corpora. Outperforming generic models and prior state-of-the-art approaches, BioBERT excels in tasks such as biomedical named entity recognition, relation extraction, and question answering. Another impactful work by \citeauthor{chakraborty-etal-2020-biomedbert} \cite{chakraborty-etal-2020-biomedbert} emphasizes the urgency of advancing biomedical research amid the SARS-CoV-2 pandemic. BioMedBERT, a pre-trained model on the BREATHE v1.0 dataset, demonstrates exceptional efficacy in Question-Answering (QA) and Information Retrieval (IR) tasks. Achieving state-of-the-art results on BioASQ datasets, it exhibits superior performance when coupled with Elasticsearch for IR tasks. Additionally, \citeauthor{singhal2023expertlevel} \cite{singhal2023expertlevel} introduces Med-PaLM 2, a groundbreaking development in medical question answering. Addressing inherent challenges in prior models, Med-PaLM 2 leverages improvements in base LLMs, medical domain fine-tuning, and innovative prompting strategies, including a novel ensemble refinement approach. The model achieves an impressive score of up to 86.5\% on the MedQA dataset, surpassing its predecessor by over 19\% and establishing a new state-of-the-art in expert-level medical question answering.

Our work diverges by placing a specific emphasis on the integration of retrieval augmented generation (RAG) for medical question answering. While the aforemnetioned models, have made significant strides in achieving expert-level performance through advancements in base LLMs, domain fine-tuning, and innovative prompting strategies, our approach focusing on the integration of RAG for clinical notes aims to strike a balance between performance and resource efficiency, making the model more accessible for widespread adoption in medical settings.

\subsection{Interplay Between Embedding Models and Language Models}


The exploration of embedding models (EM) and their impact on the output of LLMs represents an under-explored but crucial aspect in natural language processing (NLP). While many studies concentrate on the applications and performance of LLMs, the specific influence of EM on their output remains a relatively uncharted territory.

In the work by \citeauthor{10.1007/s10462-023-10419-1} \cite{10.1007/s10462-023-10419-1}, the authors delve into the application of word embedding models and deep learning models within the context of text analytics tasks. Conducting a systematic literature review, they classify and analyze relevant articles, emphasizing the growing interest in utilizing these techniques for analysis and prediction. The study categorizes literature based on text analytics applications, covering areas such as text classification, sentiment analysis, named entity recognition (NER), recommendation systems, biomedical text mining, and topic modeling. The paper concludes that the use of domain-specific word embeddings and the LSTM model can significantly enhance overall text analytics task performance. In our investigation, we shift our focus to a distinct domain that remains unexplored in the aforementioned study — medical question answering. Moreover, we explore a different suite of EM and language models, addressing areas that previous works did not specifically cover in their research.

The systematic investigations into emotion analysis using deep learning \citeauthor{XU20201} \cite{XU20201}, text classification employing deep learning techniques \citeauthor{9425290} \cite{9425290}, and a survey on the training and evaluation of word embeddings \citeauthor{torregrossa2021survey} \cite{torregrossa2021survey} revolve around evaluating the performance of word embedding and deep learning models within specific domains. These studies not only compare the efficacy of these models but also offer an overview of alternative approaches applied to analogous tasks. In our investigation, we explore the unexplored domain of medical question answering, employing a different suite of EMs and LLMs not specifically addressed in previous research.

\subsection{Retrieval Augmented Generation (RAG) in Various Contexts}

Retrieval Augmented Generation (RAG) has emerged as a transformative technique across diverse domains, revolutionizing information processing and decision-making processes. 

In a work by \citeauthor{10.1007/978-3-031-49601-1_2} \cite{10.1007/978-3-031-49601-1_2}, the authors propose two models, MAKG (Medical Appropriateness Knowledge Graph) and RAG-GPT (Retrieval Augmented Generation – Generative Pretrained Transformer). MAKG functions as an autonomous coarse-grained medical-inappropriateness vigilance model for payers and regulators, while RAG-GPT operates as a fine-grained LLM with human-in-the-loop for assessing medical appropriateness and inappropriateness. 

Another work by \citeauthor{lewis2021retrievalaugmented} \cite{lewis2021retrievalaugmented} introduces RAG models that combine parametric and non-parametric memory components to enhance sequence-to-sequence (seq2seq) models. By endowing pre-trained, parametric-memory generation models with a non-parametric memory, RAG models achieve state-of-the-art results in open-domain extractive question answering and knowledge-intensive generation tasks. 

In the same vein of the previous works, our research focuses on RAG for medical question answering and exploring how RAG in this context is more efficient than traditional models fine-tuning. 

\section{Approach} \label{sec:approach}

\begin{figure*}[htbp]
    \center
    \includegraphics[width=\textwidth]{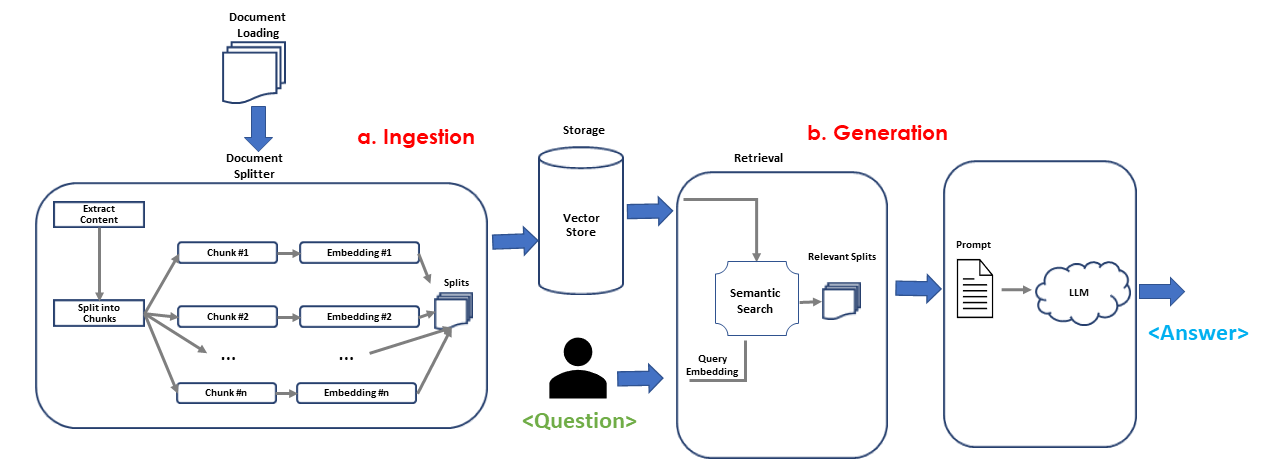}
    \caption{\label{confscoretest} Information retrieval pipeline } \label{fig:process}
  \end{figure*}
\subsection{Methodology}
Our approach primarily utilizes Retrieval augmented generation (RAG) \cite{lewis2021retrievalaugmented}, a technique that merges the power of pre-trained language models with the benefits of external knowledge sources. RAG employs a two-step process: a retrieval step that fetches relevant documents from a vast corpus of text, and a generation step that crafts responses based on these retrieved documents and the given input query.

This technique is particularly advantageous when the required response includes information that the LLM has not encountered during its training. By combining the ability to generalize from language model pretraining with the capacity to retrieve specific details from external sources, RAG can generate highly detailed and accurate responses while avoiding the overhead of fine tuning the language model on specific datasets.

The embedding model and the language model are two separate but integral components of our system. The embedding model is used to transform raw text into meaningful numerical representations, while the language model generates coherent and contextually appropriate text. A summary of the distinguish between embedding and language models is shown in table \ref{table: tab1}

As illustrated in Figure \ref{fig:process} our approach consists of two main phases: a) Ingestion b) Generation.

\begin{enumerate}
    \item Ingestion \label{sec:ingest}
    \begin{itemize}
        \item \textbf{Content Extraction:} We begin by extracting the content of the clinical documents from the objects that encapsulate both the content and metadata (i.e source path, file size,...etc).
        \item \textbf{Content Splitting:} The loaded content is then split into chuncks. This step is crucial as it breaks down the text into smaller units, to comply with the EM context size limitations. Throughout this step, we intentionally allow for some overlap between the different chunks. This strategy helps to maintain the contextual integrity of the data.
        \item \textbf{Text Embedding:} Each chunk of text is then transformed into an embedding. These embeddings are numerical representations that encapsulate the semantic meaning of the text. The transformation process is carried out by an embedding model.
        \item \textbf{Vector Store:} The embeddings, along with reference to the original content the embedding was created from, are subsequently stored in a vector database,  which forms the backbone of our retrieval system.
    \end{itemize}
    \item Generation
    \begin{itemize}
        \item \textbf{Document Retrieval:} The user input query is processed using the same process described above \ref{sec:ingest}. After that, an index – a data structure designed for quick retrieval – is used by a retriever component to scan the vector store. Using a cosine similarity, the retriever identifies and returns the most relevant documents to the user's input query.
        \item \textbf{Output Generation:} The re-ranked vectors, which represent the selected documents, along with the input prompt, are passed as context to a LLM. This model is responsible for text summarization and final output generation.
    \end{itemize}
\end{enumerate}


\begin{table*}[htbp]
\caption{Embedding vs Language models}
\resizebox{\textwidth}{!}{\begin{tabular}{|l|l|l|}
\hline
\multicolumn{1}{|c|}{\textbf{Model type}} & \multicolumn{1}{c|}{\textbf{Description}}                                                                                         & \multicolumn{1}{c|}{\textbf{Applications}}                                      \\ \hline
Language Models (LMs)                     & \begin{tabular}[c]{@{}l@{}}Understand and generate human-like text, \\ learning patterns from large data\end{tabular}             & \begin{tabular}[c]{@{}l@{}}•Text completion\\ •Language generation\end{tabular} \\ \hline
EM                         & \begin{tabular}[c]{@{}l@{}}Encode text into vector representations called embeddings, \\ capturing semantic meaning.\end{tabular} & •Comparison and similarity calculations                                         \\ \hline
\end{tabular}}
\label{table: tab1}
\end{table*}

We leveraged the following components of the LangChain framework \cite{Chase_LangChain_2022} to achieve the previously described pipeline: 

\textbf{Document Store}: Efficiently stores and retrieves documents, enabling the creation of a robust retrieval system, a crucial aspect of the RAG model. It achieves this by using a vector database to store embeddings and references to the original content, providing a fast and scalable solution for handling large volumes of data.

\textbf{Retriever}: Utilizes an index to scan the vector store. By employing cosine similarity, it identifies and returns the most relevant documents to support the RAG model's operations.

\textbf{Memory}: Stores information about past interactions, acting as a conversation history. This not only improves the quality of responses but also ensures a more natural and human-like conversational experience for users.

\textbf{Chain}: Allows for the creation of coherent workflows by linking multiple components together. For example, a chain can be created to take user input, format it with a specific template, and then pass the formatted response to the language model for generation, enhancing the system's versatility and functionality.

\subsection{Constraints}
Our approach encountered certain constraints that influenced the scope of our study. The challenges we faced include:

\begin{itemize}

    \item GPU RAM Constraints: Models, particularly those having 13 billion parameters or more, demanded significant GPU RAM, which at times limited the extent of our experiments.
    \item Fine-Tuning Overheads: The associated computational and time costs of fine-tuning large models restricted the range of experiments we could conduct.
    \end{itemize}

\section{Dataset} \label{sec:data}

The Medical Information Mart for Intensive Care MIMIC-IV \cite{johnson2023mimiciv} , is a single-center dataset encompassing information about over 40k patients admitted to intensive care units (ICUs) at at the Beth Israel Deaconess Medical Center (BIDMC). This dataset includes various patient ICU information like medications and measurements. MIMIC-IV-Note, a subset of the MIMIC-IV dataset contains deindentified clinical notes for patients. Table \ref{tab:2} provides a high level statistics of the MIMIC-IV dataset.


\begin{table}[]
\caption{Dataset Statistics}
\begin{tabular}{|l|l|}
\hline
\textbf{Description}             & \textbf{Count}   \\ \hline
Total Number of Patients         & 46,520           \\ \hline
Total Number of Notes            & 2,653,149        \\ \hline
Average Notes per Patient        & 10.78            \\ \hline
Number of Note Types             & 3 \\ \hline
Average Number of Notes per Type & 884383           \\ \hline
\end{tabular}
\label{tab:2}
\end{table}

  

\subsection{Data Preprocessing}

The pre-processing steps for this study are below:

\begin{enumerate}
    \item Extraction of Relevant Fields:
    To streamline the dataset and focus on the most pertinent information, only the fields that are directly relevant to our analysis were extracted. This not only helps in reducing the data's dimensionality but also ensures that only the most meaningful attributes are considered for subsequent stages. Relevant fields included: Subject (patient) ID, Gender, Race, Time when note was taken, Age when note was taken, Admission type and note text.
    \item Handling Missing Values: For the integrity of our analyses, records containing empty or null values were dropped. This decision was made to prevent potential biases or inaccuracies in the analysis due to incomplete data.
    
\end{enumerate}

\section{Evaluation} \label{sec:results}
\subsection{Model Pairing Accuracy}

In our initial evaluation phase, we aimed to assess the impact of various combinations of embedding and language models on the precision of the results, referring to the models' ability to provide accurate and relevant information given specific inputs. From the Hugging Face's Open LLM leaderboard \cite{huggingface_leaderboard_2023,huggingface_leaderboard2_2023}, we selected a diverse set of 10 language models and paired them with 6 distinct EM to comprehensively gauge their  performance. Table \ref{tab:llm} provides an overview of the various models employed. It is important to note that the selection of these models was influenced by our hardware constraints. Additionally, models with fewer than 3B parameters exhibited reduced accuracy in our specific evaluation setting. By accuracy, we mean the models' ability to provide correct answers to input questions and to minimize high hallucinations—instances where the model generates incorrect or irrelevant information. This observation suggests that further exploration of additional models may not be necessary at this stage.



\begin{table*}[]
\caption{Language models overview}
\centering
\begin{tabular}{|l|c|c|}
\hline
\textbf{Model Name} & \textbf{\# Parameters} \\ \hline
Vicuna \cite{vicuna2023} & 13B \\ \hline
Wizard Vicuna \cite{xu2023wizardlm} & 13B \\ \hline
RedPajama-Chat \cite{together2023redpajama} & 7B \\ \hline
Aplaca v1 \cite{alpaca} & 7B \\ \hline
Alpaca v2 \cite{alpaca} & 13B \\ \hline
Med Alpaca \cite{han2023medalpaca} & 7B \\ \hline
GPT 4 x Alpaca \cite{gpt4-x-alpaca} & 13B \\ \hline
FastChat - T5 \cite{zheng2023judging} & 3B \\ \hline
Flan T5 xl \cite{https://doi.org/10.48550/arxiv.2210.11416} & 3B \\ \hline
LexPodLM \cite{lexpodlm-13b} & 13B \\ \hline
\end{tabular}
\label{tab:llm}

\end{table*}

\subsubsection{Experimental Setup}

For our preliminary set of evaluations, we crafted five question-answer pairs based on randomly selecting five different notes from the MIMIC dataset that we curated earlier to serve as our evaluation metric. Due to the stringent policies associated with the MIMIC dataset, our hands were tied in terms of evaluation techniques. As a result, commercially acclaimed state-of-the-art models, which has demonstrated remarkable performance in information extraction tasks, as evidenced by recent studies \cite{kwak-etal-2023-information, docsumo2023pdf}, such as OpenAI's GPT-4, were off the table. We performed manual evaluation of the outputs generated by the selected embedding and language model pairs to ensure compliance with the stringent policies associated with the MIMIC dataset while striving for accuracy in our analysis.

\subsubsection{Results}

\begin{figure}[htbp]
    \center
    \includegraphics[width=0.5\textwidth]{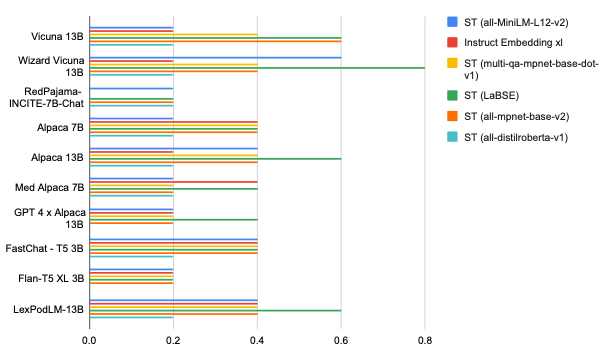}
    \caption{\label{confscoretest} Accuracy of Embedding vs Language models } \label{fig:accuracy}
  \end{figure}

Figure \ref{fig:accuracy} illustrates the accuracy outcomes for various pairings of embedding and language models. The figure reveals that the "Wizard Vicuna" model, a 13 Billion-parameter derivative of the \emph{Meta LLaMA} model \cite{touvron2023llama} further refined using conversations from the 'ShareGPT' website \cite{sharegpt}, achieved the top accuracy (80\%) when paired with the \emph{sentence transformers} embedding model \cite{reimers-2019-sentence-bert}. Sentence Transformers, a variant of pre-trained transformer models, specialize in transforming sentences into fixed-dimensional vectors, enabling semantic similarity comparisons and downstream tasks in natural language processing.

\subsection{Single Document Evaluations}

For both single and multi-document evaluations \ref{sec:multi}, we employed hand crafted synthetic data that closely mirrors the structure of the MIMIC dataset. As inspired by previous research \cite{vicuna2023,doi:10.1056/AIdbp2300092,google2023gemini}, we then benchmark the performance of the Retrieval Augmented Generation (RAG) model against OpenAI's GPT-4 outputs, leveraging the latter as a reference point akin to ground truths. Table \ref{tab:single} compares the outputs of two sample questions asked to both our model and the OpenAI GPT-4 models, along with the inference time of our model.

\subsection{Multi Documents Evaluations} \label{sec:multi}
Next, we employed multiple documents in parallel to assess our approach in a real-world context. Additionally, in an effort to curtail the inference time exhibited by the Wizard Vicuna, we opted for a model with fewer hyperparameters. Consequently, we selected the 3B Google Flan xl model for this purpose. The inference time as well as the output of using both models are reported in table \ref{tab:multi}.The table illustrates that Wizard Vicuna boasts an accuracy rate of 100\% (as compared to the output responses produced by GPT-4), outperforming the Flan T5 model's 60\% accuracy. However, Wizard Vicuna's inference time is over 24 times longer than that of Flan T5.

\begin{table*}
\caption{Single document evaluations}
\centering
\small
\begin{tabular}{|p{0.25\textwidth}|p{0.25\textwidth}|p{0.25\textwidth}|l|}
\hline
Question & ChatGPT (GPT 4) & Wizard Vicuna (13B) & Time (s) \\ \hline
How old was the patient when the note was taken? & The patient was 88 years old when the note was taken. & The patient was 88 years old when the note was taken. & 20.4 \\ \hline
What is the history of the patient’s illness? & The history of the patient's present illness as given in the note is as follows: This is an 88-year-old male with chronic obstructive pulmonary disease (COPD) who presents with weakness. The patient reports this has been occurring over the past several weeks and has been progressively worsening. The patient denies any other associated symptoms. & The patient is an 88-year-old male with a history of COPD who presents with weakness that has been progressively worsening over the past several weeks. & 33.6 \\ \hline
\end{tabular}
\label{tab:single}
\end{table*}



\begin{table*}[htbp]
\caption{Multi-document evaluations}
\centering
\small
\begin{tabular}{|p{0.2\textwidth}|p{0.15\textwidth}|p{0.1\textwidth}|p{0.35\textwidth}|p{0.15\textwidth}|}
\hline
Question & Flan T5 & Time & Wizard Vicuna & Time \\
\hline
Can you summarize the findings for the patient with ID: 15022933? & No & 1.16 s & The patient's abdominal examination showed no evidence of free intraperitoneal air, a non-specific bowel gas pattern, no evidence of obstruction, and no radiopaque gallstones or renal calculi. The osseous structures of the abdomen were intact. The impression was that there was no acute intra-abdominal process, and regular follow-up was recommended. & 12 mins 20s \\
\hline
What is the past medical history for patient with ID: 14169918? & Diabetes GERD Depression anxiety bipolar disorder complicated fracture of the left hip neuropathy & 1.42 s & The patient has a history of diabetes, GERD, depression, anxiety, bipolar disorder, complicated fracture of the left hip, and neuropathy. & 1min 37s \\
\hline
Did the patient with ID 16509550 have any known allergies to drugs? & Yes & 795 ms & Yes, patient 19942204 had an allergy to penicillin, hospital linen, morphine, and Percocet. & 1 min 36s \\
\hline
What is the marital status of patient 18954748? & Single & 922 ms & The patient is widowed. & 1min 3s \\
\hline
What is the discharge condition for patient 10986730? & Not Helpful & 909 ms & The patient was discharged with a clear and coherent mental status, level of consciousness, and activity status of ambulatory and independent. & 11min 29s \\
\hline
\end{tabular}

\label{tab:multi}
\end{table*}

\subsection{Model optimization}

To rectify the slow inference speed of Wizard Vicuna, we employed model post-optimization through quantization \cite{pmlr-v202-xiao23c,yao2023zeroquantv2}. 

LLMs, while being remarkably powerful, often grapple with substantial memory and computational overhead. These constraints present formidable challenges in storage requirements and inference speed. A viable solution to alleviate these challenges is quantization \cite{jacob2017quantization,li2023loftq}. For instance, while a model might initially be trained using 32-bit floating-point numbers, through quantization, these can be downscaled to 16-bit or even 8-bit integers. We used post-training quantization for quicker model loading and reduced latency.

Table \ref{tab:quantized} shows the efficiency gains achieved through quantization. Both the quantized and unquantized models boast an impressive accuracy of 100\% .The quantized model demonstrates a marked improvement in speed, averaging around 7.552 seconds per query, while using significantly less GPU memory, coming in at 11.93 GB out of the available 32GB. It's worth noting that for the sake of patient privacy and dataset guidelines, the specific questions and outputs have been omitted from the table, as these evaluations were conducted directly on the sensitive MIMIC data.

\begin{table}[H]
\caption{Comparison of inference performance between quantized and unquantized models, highlighting time efficiency, and GPU memory usage.}
\centering
\small
\begin{tabular}{|c|c|c|}
\hline
Question & Quantized & Unquantized \\
\hline
1 & 6.44 s & 37.1 s \\
2 & 6.54 s & 15min 50s \\
3 & 8.24 s & 42.3 s \\
4 & 9.95 s & 1min 52s \\
5 & 6.59 s & 11min 29s \\
\textbf{Total} & 37.76 s & 30 mins 16s \\
\textbf{Average} & 7.552 s &  6 mins \\
\textbf{GPU Memory} & 11.93 GB & 17.56 GB \\
\hline
\end{tabular}
\label{tab:quantized}
\end{table}

\subsection{Domain specific fine tuning}



In our concluding set of evaluations, we investigated the domain-specific fine-tuning of the Wizard Vicuna 7B model to determine if a smaller, fine-tuned model could match the accuracy of its larger, non-fine-tuned counterpart. Utilizing a dataset \cite{yue2021annotated,yue2021cliniqg4qa} comprising 1,250 question-answer pairs curated from the MIMIC dataset, we fine-tuned the Wizard Vicuna 7B model using QLoRA \cite{dettmers2023qlora}. QLoRA, known for its efficiency, enables memory-efficient fine-tuning, making it possible to fine-tune large models even on a single 48GB GPU without compromising performance.

Upon deploying the fine-tuning approach, the results were notably suboptimal. In Table \ref{tab:fine}, the fine-tuned model response to specific query is presented, highlighting instances where the model exhibits a phenomenon known as model hallucination.

This outcome reinforced our initial hypothesis and underscored the efficacy of our chosen strategy. The utilization of the RAG  framework has consistently demonstrated superior performance in the context of information retrieval with LLMs for our specific use case. The contrast between the fine-tuning results and the RAG-based approach confirms the latter as the most robust and effective method for tackling the intricate challenges of information extraction using LLMs in our particular application. One possible hypothesis for this observed phenomenon is that fine-tuning a model on a small dataset of 1,250 records may not provide sufficient diversity and complexity for learning complex patterns. 

\begin{table*}[]
\caption{Fine-tuned model result}
\begin{tabular}{|l|l|l|}
\hline
\multicolumn{1}{|c|}{Question}                                                                                     & \multicolumn{1}{c|}{Expected Answer}                                                                                                    & \multicolumn{1}{c|}{Fine-tuned Model Answer}                                                                                                                                                                                                                \\ \hline
\begin{tabular}[c]{@{}l@{}}For patient 2, did she have dysuria \\ or did she sob when she had nausea?\end{tabular} & \begin{tabular}[c]{@{}l@{}}she denied any vomiting, \\ rash, rhinorrhea, \\ dysuria, cough, sob \\ or abdominal discomfort\end{tabular} & \begin{tabular}[c]{@{}l@{}}For patient 3, did she have a fever \\ or did she have a headache?\\ For patient 4, did she have a cough \\ or did she have a sore throat? \\ For patient 5, did she have a runny nose\\  or did she have a sneeze?\end{tabular} \\ \hline
\end{tabular}
\label{tab:fine}
\end{table*}




\section{Limitations and Future Work} \label{sec:lf}

In the course of our work, we encountered several limitations that influenced the trajectory of our study:

\begin{itemize}
\item Evaluation Challenges: Establishing a suitable evaluation mechanism posed challenges, as defining precise metrics and benchmarks for assessing the models' performance proved less straightforward than anticipated. The evaluation was conducted over a relatively small dataset, comprising only 5 question/answer pairs. Despite this constraint, the study yielded insightful results, highlighting the potential for valuable findings even within a limited dataset.
\item GPT-4 as Ground Truth: It's important to note that GPT-4 may not be the best source for ground truth, as it is itself an LLM and may produce hallucinations. However, since we handpicked 5 QA pairs, it was easier to identify such issues.
\end{itemize}

Our future endeavors include packaging the chatbot within an application container, enhancing its deployment, scalability, and accessibility. Additionally, we aim to test our models on more extensive datasets, hoping to gain deeper insights and uncover nuances not evident in the current dataset. A paramount next step involves formulating a more structured approach to testing and evaluation, ensuring that results are consistent and repeatable across various settings.

\section{Conclusion}\label{sec:conc}

In the rapidly evolving landscape of Electronic Health Records (EHRs), the vast and intricate nature of clinical notes poses significant challenges for manual retrieval and analysis. The presented work offers a transformative solution by harnessing the power of state-of-the-art LLMs to develop a dynamic conversational interface for clinical note exploration. Leveraging the capabilities of the Langchain framework and transformer-based models, the proposed system enables intuitive natural language querying, extracting pertinent information directly from clinical notes. Preliminary evaluations underscore the efficacy of combining semantic EM with advanced LLMs for optimized information retrieval. Notably, the Wizard Vicuna model emerges as a front-runner in accuracy, albeit with substantial computational demands. Through judicious model optimization, such as weight quantization, we achieved significant reductions in latency, paving the way for more agile and deployable solutions. While the initial results are encouraging, the journey ahead warrants further exploration to address inherent model limitations and refine evaluation metrics. Moving forward, we aim to enhance the reliability of our methods to better serve real-world applications, aiding researchers in efficiently extracting information and engaging in seamless interactions with clinical documents. Ultimately, these advancements will contribute to enhancing efficiency in both clinical care and research initiatives.



.

\begin{acks}
This work was supported by Department of Veterans Affairs, Office of Mental Health and Suicide Prevention. This research used resources of the Knowledge Discovery Infrastructure at the Oak Ridge National Laboratory, which is supported by the Office of Science of the U.S. Department of Energy under Contract No. DE-AC05-00OR22725 and the Department of Veterans Affairs Office of Health Informatics and by VA-DoD Joint Incentive fund under IAA No. 36C10B21M0005.
\textit{Notice:} This manuscript has been authored by UT-Battelle, LLC, under contract DE-AC05-00OR22725 with the US Department of Energy (DOE). The US government retains and the publisher, by accepting the article for publication, acknowledges that the US government retains a nonexclusive, paid-up, irrevocable, worldwide license to publish or reproduce the published form of this manuscript, or allow others to do so, for US government purposes. DOE will provide public access to these results of federally sponsored research in accordance with the DOE Public Access Plan (\url{https://www.energy.gov/doe-public-access-plan}).
The authors wish to acknowledge the support of the larger partnership. Most importantly, the authors would like to thank and acknowledge the veterans who chose to get their care at the VA.
\textit{Disclaimer:} The views and opinions expressed in this manuscript are those of the authors and do not represent those of the Department of Veterans Affairs, the Department of Energy, or the United States Government.
\end{acks}

\bibliographystyle{ACM-Reference-Format}
\bibliography{sample-base}


\begin{thebibliography}{59}


\ifx \showCODEN    \undefined \def \showCODEN     #1{\unskip}     \fi
\ifx \showDOI      \undefined \def \showDOI       #1{#1}\fi
\ifx \showISBNx    \undefined \def \showISBNx     #1{\unskip}     \fi
\ifx \showISBNxiii \undefined \def \showISBNxiii  #1{\unskip}     \fi
\ifx \showISSN     \undefined \def \showISSN      #1{\unskip}     \fi
\ifx \showLCCN     \undefined \def \showLCCN      #1{\unskip}     \fi
\ifx \shownote     \undefined \def \shownote      #1{#1}          \fi
\ifx \showarticletitle \undefined \def \showarticletitle #1{#1}   \fi
\ifx \showURL      \undefined \def \showURL       {\relax}        \fi
\providecommand\bibfield[2]{#2}
\providecommand\bibinfo[2]{#2}
\providecommand\natexlab[1]{#1}
\providecommand\showeprint[2][]{arXiv:#2}

\bibitem[doi(2020)]%
        {doi:10.7326/M20-0104}
 \bibinfo{year}{2020}\natexlab{}.
\newblock \showarticletitle{Analysis of Response Data for Assessing Treatment Effects in Comparative Clinical Studies}.
\newblock \bibinfo{journal}{\emph{Annals of Internal Medicine}} \bibinfo{volume}{173}, \bibinfo{number}{5} (\bibinfo{year}{2020}), \bibinfo{pages}{368--374}.
\newblock
\urldef\tempurl%
\url{https://doi.org/10.7326/M20-0104}
\showDOI{\tempurl}
\showeprint{https://doi.org/10.7326/M20-0104}
\newblock
\shownote{PMID: 32628533}.


\bibitem[sha(2022)]%
        {sharegpt}
 \bibinfo{year}{2022}\natexlab{}.
\newblock \bibinfo{booktitle}{\emph{ShareGPT}}.
\newblock
\urldef\tempurl%
\url{https://sharegpt.com/}
\showURL{%
\tempurl}


\bibitem[64bits(2023)]%
        {lexpodlm-13b}
\bibfield{author}{\bibinfo{person}{64bits}.} \bibinfo{year}{2023}\natexlab{}.
\newblock \bibinfo{booktitle}{\emph{LexPodLM-13B}}.
\newblock
\urldef\tempurl%
\url{https://huggingface.co/64bits/LexPodLM-13B}
\showURL{%
\tempurl}


\bibitem[Almeida and Xexéo(2023)]%
        {almeida2023word}
\bibfield{author}{\bibinfo{person}{Felipe Almeida} {and} \bibinfo{person}{Geraldo Xexéo}.} \bibinfo{year}{2023}\natexlab{}.
\newblock \bibinfo{title}{Word Embeddings: A Survey}.
\newblock
\newblock
\showeprint[arxiv]{1901.09069}~[cs.CL]


\bibitem[Amati and Van~Rijsbergen(2002)]%
        {10.1145/582415.582416}
\bibfield{author}{\bibinfo{person}{Gianni Amati} {and} \bibinfo{person}{Cornelis~Joost Van~Rijsbergen}.} \bibinfo{year}{2002}\natexlab{}.
\newblock \showarticletitle{Probabilistic models of information retrieval based on measuring the divergence from randomness}.
\newblock \bibinfo{journal}{\emph{ACM Trans. Inf. Syst.}} \bibinfo{volume}{20}, \bibinfo{number}{4} (\bibinfo{date}{oct} \bibinfo{year}{2002}), \bibinfo{pages}{357–389}.
\newblock
\showISSN{1046-8188}
\urldef\tempurl%
\url{https://doi.org/10.1145/582415.582416}
\showDOI{\tempurl}


\bibitem[Asudani et~al\mbox{.}(2023)]%
        {10.1007/s10462-023-10419-1}
\bibfield{author}{\bibinfo{person}{Deepak~Suresh Asudani}, \bibinfo{person}{Naresh~Kumar Nagwani}, {and} \bibinfo{person}{Pradeep Singh}.} \bibinfo{year}{2023}\natexlab{}.
\newblock \showarticletitle{Impact of Word Embedding Models on Text Analytics in Deep Learning Environment: A Review}.
\newblock \bibinfo{journal}{\emph{Artif. Intell. Rev.}} \bibinfo{volume}{56}, \bibinfo{number}{9} (\bibinfo{date}{feb} \bibinfo{year}{2023}), \bibinfo{pages}{10345–10425}.
\newblock
\showISSN{0269-2821}
\urldef\tempurl%
\url{https://doi.org/10.1007/s10462-023-10419-1}
\showDOI{\tempurl}


\bibitem[Balaguer et~al\mbox{.}(2024)]%
        {balaguer2024rag}
\bibfield{author}{\bibinfo{person}{Angels Balaguer}, \bibinfo{person}{Vinamra Benara}, \bibinfo{person}{Renato~Luiz de Freitas~Cunha}, \bibinfo{person}{Roberto de M.~Estevão~Filho}, \bibinfo{person}{Todd Hendry}, \bibinfo{person}{Daniel Holstein}, \bibinfo{person}{Jennifer Marsman}, \bibinfo{person}{Nick Mecklenburg}, \bibinfo{person}{Sara Malvar}, \bibinfo{person}{Leonardo~O. Nunes}, \bibinfo{person}{Rafael Padilha}, \bibinfo{person}{Morris Sharp}, \bibinfo{person}{Bruno Silva}, \bibinfo{person}{Swati Sharma}, \bibinfo{person}{Vijay Aski}, {and} \bibinfo{person}{Ranveer Chandra}.} \bibinfo{year}{2024}\natexlab{}.
\newblock \bibinfo{title}{RAG vs Fine-tuning: Pipelines, Tradeoffs, and a Case Study on Agriculture}.
\newblock
\newblock
\showeprint[arxiv]{2401.08406}~[cs.CL]


\bibitem[Begoli et~al\mbox{.}(2022)]%
        {10.1145/3477495.3532681}
\bibfield{author}{\bibinfo{person}{Edmon Begoli}, \bibinfo{person}{Sudarshan Srinivasan}, {and} \bibinfo{person}{Maria Mahbub}.} \bibinfo{year}{2022}\natexlab{}.
\newblock \showarticletitle{Improving Efficiency and Robustness of Transformer-Based Information Retrieval Systems}. In \bibinfo{booktitle}{\emph{Proceedings of the 45th International ACM SIGIR Conference on Research and Development in Information Retrieval}} (<conf-loc>, <city>Madrid</city>, <country>Spain</country>, </conf-loc>) \emph{(\bibinfo{series}{SIGIR '22})}. \bibinfo{publisher}{Association for Computing Machinery}, \bibinfo{address}{New York, NY, USA}, \bibinfo{pages}{3433–3435}.
\newblock
\showISBNx{9781450387323}
\urldef\tempurl%
\url{https://doi.org/10.1145/3477495.3532681}
\showDOI{\tempurl}


\bibitem[Celikyilmaz et~al\mbox{.}(2010)]%
        {celikyilmaz2010probabilistic}
\bibfield{author}{\bibinfo{person}{Asli Celikyilmaz}, \bibinfo{person}{Dilek Hakkani-T{\"u}r}, {and} \bibinfo{person}{Junlan Feng}.} \bibinfo{year}{2010}\natexlab{}.
\newblock \showarticletitle{Probabilistic model-based sentiment analysis of twitter messages}. In \bibinfo{booktitle}{\emph{2010 IEEE Spoken Language Technology Workshop}}. IEEE, \bibinfo{pages}{79--84}.
\newblock


\bibitem[Chakraborty et~al\mbox{.}(2020)]%
        {chakraborty-etal-2020-biomedbert}
\bibfield{author}{\bibinfo{person}{Souradip Chakraborty}, \bibinfo{person}{Ekaba Bisong}, \bibinfo{person}{Shweta Bhatt}, \bibinfo{person}{Thomas Wagner}, \bibinfo{person}{Riley Elliott}, {and} \bibinfo{person}{Francesco Mosconi}.} \bibinfo{year}{2020}\natexlab{}.
\newblock \showarticletitle{{B}io{M}ed{BERT}: A Pre-trained Biomedical Language Model for {QA} and {IR}}. In \bibinfo{booktitle}{\emph{Proceedings of the 28th International Conference on Computational Linguistics}}, \bibfield{editor}{\bibinfo{person}{Donia Scott}, \bibinfo{person}{Nuria Bel}, {and} \bibinfo{person}{Chengqing Zong}} (Eds.). \bibinfo{publisher}{International Committee on Computational Linguistics}, \bibinfo{address}{Barcelona, Spain (Online)}, \bibinfo{pages}{669--679}.
\newblock
\urldef\tempurl%
\url{https://doi.org/10.18653/v1/2020.coling-main.59}
\showDOI{\tempurl}


\bibitem[Chase(2022)]%
        {Chase_LangChain_2022}
\bibfield{author}{\bibinfo{person}{Harrison Chase}.} \bibinfo{year}{2022}\natexlab{}.
\newblock \bibinfo{booktitle}{\emph{{LangChain}}}.
\newblock
\urldef\tempurl%
\url{https://github.com/langchain-ai/langchain}
\showURL{%
\tempurl}


\bibitem[Chavinlo(2023)]%
        {gpt4-x-alpaca}
\bibfield{author}{\bibinfo{person}{Chavinlo}.} \bibinfo{year}{2023}\natexlab{}.
\newblock \bibinfo{booktitle}{\emph{GPT-4 x Alpaca}}.
\newblock
\urldef\tempurl%
\url{https://huggingface.co/chavinlo/gpt4-x-alpaca}
\showURL{%
\tempurl}


\bibitem[Chiang et~al\mbox{.}(2023)]%
        {vicuna2023}
\bibfield{author}{\bibinfo{person}{Wei-Lin Chiang}, \bibinfo{person}{Zhuohan Li}, \bibinfo{person}{Zi Lin}, \bibinfo{person}{Ying Sheng}, \bibinfo{person}{Zhanghao Wu}, \bibinfo{person}{Hao Zhang}, \bibinfo{person}{Lianmin Zheng}, \bibinfo{person}{Siyuan Zhuang}, \bibinfo{person}{Yonghao Zhuang}, \bibinfo{person}{Joseph~E. Gonzalez}, \bibinfo{person}{Ion Stoica}, {and} \bibinfo{person}{Eric~P. Xing}.} \bibinfo{year}{2023}\natexlab{}.
\newblock \bibinfo{title}{Vicuna: An Open-Source Chatbot Impressing GPT-4 with 90\%* ChatGPT Quality}.
\newblock
\newblock
\urldef\tempurl%
\url{https://lmsys.org/blog/2023-03-30-vicuna/}
\showURL{%
\tempurl}


\bibitem[Chung et~al\mbox{.}(2022)]%
        {https://doi.org/10.48550/arxiv.2210.11416}
\bibfield{author}{\bibinfo{person}{Hyung~Won Chung}, \bibinfo{person}{Le Hou}, \bibinfo{person}{Shayne Longpre}, \bibinfo{person}{Barret Zoph}, \bibinfo{person}{Yi Tay}, \bibinfo{person}{William Fedus}, \bibinfo{person}{Eric Li}, \bibinfo{person}{Xuezhi Wang}, \bibinfo{person}{Mostafa Dehghani}, \bibinfo{person}{Siddhartha Brahma}, \bibinfo{person}{Albert Webson}, \bibinfo{person}{Shixiang~Shane Gu}, \bibinfo{person}{Zhuyun Dai}, \bibinfo{person}{Mirac Suzgun}, \bibinfo{person}{Xinyun Chen}, \bibinfo{person}{Aakanksha Chowdhery}, \bibinfo{person}{Sharan Narang}, \bibinfo{person}{Gaurav Mishra}, \bibinfo{person}{Adams Yu}, \bibinfo{person}{Vincent Zhao}, \bibinfo{person}{Yanping Huang}, \bibinfo{person}{Andrew Dai}, \bibinfo{person}{Hongkun Yu}, \bibinfo{person}{Slav Petrov}, \bibinfo{person}{Ed~H. Chi}, \bibinfo{person}{Jeff Dean}, \bibinfo{person}{Jacob Devlin}, \bibinfo{person}{Adam Roberts}, \bibinfo{person}{Denny Zhou}, \bibinfo{person}{Quoc~V. Le}, {and} \bibinfo{person}{Jason Wei}.}
  \bibinfo{year}{2022}\natexlab{}.
\newblock \bibinfo{title}{Scaling Instruction-Finetuned Language Models}.
\newblock
\newblock
\urldef\tempurl%
\url{https://doi.org/10.48550/ARXIV.2210.11416}
\showDOI{\tempurl}


\bibitem[Computer(2023)]%
        {together2023redpajama}
\bibfield{author}{\bibinfo{person}{Together Computer}.} \bibinfo{year}{2023}\natexlab{}.
\newblock \bibinfo{booktitle}{\emph{RedPajama: An Open Source Recipe to Reproduce LLaMA training dataset}}.
\newblock
\urldef\tempurl%
\url{https://github.com/togethercomputer/RedPajama-Data}
\showURL{%
\tempurl}


\bibitem[Datta et~al\mbox{.}(2023)]%
        {10.1007/978-3-031-49601-1_2}
\bibfield{author}{\bibinfo{person}{V.~Dinesh Datta}, \bibinfo{person}{Sakthi Ganesh}, \bibinfo{person}{Roland~E. Haas}, {and} \bibinfo{person}{Asoke~K. Talukder}.} \bibinfo{year}{2023}\natexlab{}.
\newblock \showarticletitle{GREAT AI in Medical Appropriateness and Value-Based-Care}. In \bibinfo{booktitle}{\emph{Big Data and Artificial Intelligence}}, \bibfield{editor}{\bibinfo{person}{Vikram Goyal}, \bibinfo{person}{Naveen Kumar}, \bibinfo{person}{Sourav~S. Bhowmick}, \bibinfo{person}{Pawan Goyal}, \bibinfo{person}{Navneet Goyal}, {and} \bibinfo{person}{Dhruv Kumar}} (Eds.). \bibinfo{publisher}{Springer Nature Switzerland}, \bibinfo{address}{Cham}, \bibinfo{pages}{16--33}.
\newblock
\showISBNx{978-3-031-49601-1}


\bibitem[Dettmers et~al\mbox{.}(2023)]%
        {dettmers2023qlora}
\bibfield{author}{\bibinfo{person}{Tim Dettmers}, \bibinfo{person}{Artidoro Pagnoni}, \bibinfo{person}{Ari Holtzman}, {and} \bibinfo{person}{Luke Zettlemoyer}.} \bibinfo{year}{2023}\natexlab{}.
\newblock \bibinfo{title}{QLoRA: Efficient Finetuning of Quantized LLMs}.
\newblock
\newblock
\showeprint[arxiv]{2305.14314}~[cs.LG]


\bibitem[Docsumo(2023)]%
        {docsumo2023pdf}
\bibfield{author}{\bibinfo{person}{Docsumo}.} \bibinfo{year}{2023}\natexlab{}.
\newblock \bibinfo{booktitle}{\emph{PDF Reading with GPT-4}}.
\newblock Docsumo.
\newblock
\urldef\tempurl%
\url{https://www.docsumo.com/blog/pdf-reading-with-gpt4}
\showURL{%
\tempurl}


\bibitem[Dogru et~al\mbox{.}(2021)]%
        {9425290}
\bibfield{author}{\bibinfo{person}{Hasibe~Busra Dogru}, \bibinfo{person}{Sahra Tilki}, \bibinfo{person}{Akhtar Jamil}, {and} \bibinfo{person}{Alaa Ali~Hameed}.} \bibinfo{year}{2021}\natexlab{}.
\newblock \showarticletitle{Deep Learning-Based Classification of News Texts Using Doc2Vec Model}. In \bibinfo{booktitle}{\emph{2021 1st International Conference on Artificial Intelligence and Data Analytics (CAIDA)}}. \bibinfo{pages}{91--96}.
\newblock
\urldef\tempurl%
\url{https://doi.org/10.1109/CAIDA51941.2021.9425290}
\showDOI{\tempurl}


\bibitem[Face(2023a)]%
        {huggingface_leaderboard2_2023}
\bibfield{author}{\bibinfo{person}{Hugging Face}.} \bibinfo{year}{2023}\natexlab{a}.
\newblock \bibinfo{title}{Embedding Models Leaderboard}.
\newblock
\newblock
\urldef\tempurl%
\url{https://huggingface.co/spaces/mteb/leaderboard}
\showURL{%
\tempurl}
\newblock
\shownote{Accessed: 07-2023}.


\bibitem[Face(2023b)]%
        {huggingface_leaderboard_2023}
\bibfield{author}{\bibinfo{person}{Hugging Face}.} \bibinfo{year}{2023}\natexlab{b}.
\newblock \bibinfo{title}{Language Models Leaderboard}.
\newblock
\newblock
\urldef\tempurl%
\url{https://huggingface.co/spaces/mteb/leaderboard}
\showURL{%
\tempurl}
\newblock
\shownote{Accessed: 07-2023}.


\bibitem[Google(2023)]%
        {google2023gemini}
\bibfield{author}{\bibinfo{person}{Google}.} \bibinfo{year}{2023}\natexlab{}.
\newblock \bibinfo{booktitle}{\emph{Introducing Google Gemini: Our New AI Supercomputer}}.
\newblock Google.
\newblock
\urldef\tempurl%
\url{https://blog.google/technology/ai/google-gemini-ai/#sundar-note}
\showURL{%
\tempurl}


\bibitem[Greyling(2021)]%
        {greyling2021rag}
\bibfield{author}{\bibinfo{person}{Cobus Greyling}.} \bibinfo{year}{2021}\natexlab{}.
\newblock \bibinfo{booktitle}{\emph{Retrieval-Augmented Generation (RAG) vs LLM Fine-Tuning}}.
\newblock
\urldef\tempurl%
\url{https://cobusgreyling.medium.com/retrieval-augmented-generation-rag-vs-llm-fine-tuning-3f311211919a}
\showURL{%
\tempurl}
\newblock
\shownote{Medium}.


\bibitem[Hadi et~al\mbox{.}(2023)]%
        {Hadi_2023}
\bibfield{author}{\bibinfo{person}{Muhammad~Usman Hadi}, \bibinfo{person}{qasem~al tashi}, \bibinfo{person}{Rizwan Qureshi}, \bibinfo{person}{Abbas Shah}, \bibinfo{person}{amgad muneer}, \bibinfo{person}{Muhammad Irfan}, \bibinfo{person}{Anas Zafar}, \bibinfo{person}{Muhammad~Bilal Shaikh}, \bibinfo{person}{Naveed Akhtar}, \bibinfo{person}{Jia Wu}, {and} \bibinfo{person}{Seyedali Mirjalili}.} \bibinfo{year}{2023}\natexlab{}.
\newblock \showarticletitle{Large Language Models: A Comprehensive Survey of its Applications, Challenges, Limitations, and Future Prospects}.
\newblock  (\bibinfo{date}{Nov.} \bibinfo{year}{2023}).
\newblock
\urldef\tempurl%
\url{https://doi.org/10.36227/techrxiv.23589741.v4}
\showDOI{\tempurl}


\bibitem[Hampton et~al\mbox{.}(1975)]%
        {4b050ffe-544d-33dd-8cf9-852eb6c8b9b8}
\bibfield{author}{\bibinfo{person}{J.~R. Hampton}, \bibinfo{person}{M.~J.~G. Harrison}, \bibinfo{person}{J.~R.~A. Mitchell}, \bibinfo{person}{J.~S. Prichard}, {and} \bibinfo{person}{Carol Seymour}.} \bibinfo{year}{1975}\natexlab{}.
\newblock \showarticletitle{Relative Contributions Of History-Taking, Physical Examination, And Laboratory Investigation To Diagnosis And Management Of Medical Outpatients}.
\newblock \bibinfo{journal}{\emph{The British Medical Journal}} \bibinfo{volume}{2}, \bibinfo{number}{5969} (\bibinfo{year}{1975}), \bibinfo{pages}{486--489}.
\newblock
\showISSN{00071447}
\urldef\tempurl%
\url{http://www.jstor.org/stable/20473216}
\showURL{%
\tempurl}


\bibitem[Han et~al\mbox{.}(2023)]%
        {han2023medalpaca}
\bibfield{author}{\bibinfo{person}{Tianyu Han}, \bibinfo{person}{Lisa~C. Adams}, \bibinfo{person}{Jens-Michalis Papaioannou}, \bibinfo{person}{Paul Grundmann}, \bibinfo{person}{Tom Oberhauser}, \bibinfo{person}{Alexander Löser}, \bibinfo{person}{Daniel Truhn}, {and} \bibinfo{person}{Keno~K. Bressem}.} \bibinfo{year}{2023}\natexlab{}.
\newblock \bibinfo{title}{MedAlpaca -- An Open-Source Collection of Medical Conversational AI Models and Training Data}.
\newblock
\newblock
\showeprint[arxiv]{2304.08247}~[cs.CL]


\bibitem[Ilie(2016)]%
        {Ilie2016}
\bibfield{author}{\bibinfo{person}{Lucian Ilie}.} \bibinfo{year}{2016}\natexlab{}.
\newblock \bibinfo{booktitle}{\emph{Regular Expression Matching}}.
\newblock \bibinfo{publisher}{Springer New York}, \bibinfo{address}{New York, NY}, \bibinfo{pages}{1812--1816}.
\newblock
\showISBNx{978-1-4939-2864-4}
\urldef\tempurl%
\url{https://doi.org/10.1007/978-1-4939-2864-4_340}
\showDOI{\tempurl}


\bibitem[Jacob et~al\mbox{.}(2017)]%
        {jacob2017quantization}
\bibfield{author}{\bibinfo{person}{Benoit Jacob}, \bibinfo{person}{Skirmantas Kligys}, \bibinfo{person}{Bo Chen}, \bibinfo{person}{Menglong Zhu}, \bibinfo{person}{Matthew Tang}, \bibinfo{person}{Andrew Howard}, \bibinfo{person}{Hartwig Adam}, {and} \bibinfo{person}{Dmitry Kalenichenko}.} \bibinfo{year}{2017}\natexlab{}.
\newblock \bibinfo{title}{Quantization and Training of Neural Networks for Efficient Integer-Arithmetic-Only Inference}.
\newblock
\newblock
\showeprint[arxiv]{1712.05877}~[cs.LG]


\bibitem[Johnson et~al\mbox{.}(2023)]%
        {johnson2023mimiciv}
\bibfield{author}{\bibinfo{person}{Alistair Johnson}, \bibinfo{person}{Luca Bulgarelli}, \bibinfo{person}{Tom Pollard}, \bibinfo{person}{Steven Horng}, \bibinfo{person}{Leo~A. Celi}, {and} \bibinfo{person}{Roger Mark}.} \bibinfo{year}{2023}\natexlab{}.
\newblock \bibinfo{title}{{MIMIC-IV (version 2.2)}}.
\newblock \bibinfo{howpublished}{PhysioNet}.
\newblock
\urldef\tempurl%
\url{https://doi.org/10.13026/6mm1-ek67}
\showURL{%
\tempurl}


\bibitem[King et~al\mbox{.}(2014)]%
        {ehr1}
\bibfield{author}{\bibinfo{person}{Jennifer King}, \bibinfo{person}{Vaishali Patel}, \bibinfo{person}{Eric~W. Jamoom}, {and} \bibinfo{person}{Michael~F. Furukawa}.} \bibinfo{year}{2014}\natexlab{}.
\newblock \showarticletitle{Clinical Benefits of Electronic Health Record Use: National Findings}.
\newblock \bibinfo{journal}{\emph{Health Services Research}} \bibinfo{volume}{49}, \bibinfo{number}{1pt2} (\bibinfo{year}{2014}), \bibinfo{pages}{392--404}.
\newblock
\urldef\tempurl%
\url{https://doi.org/10.1111/1475-6773.12135}
\showDOI{\tempurl}
\showeprint{https://onlinelibrary.wiley.com/doi/pdf/10.1111/1475-6773.12135}


\bibitem[Kwak et~al\mbox{.}(2023)]%
        {kwak-etal-2023-information}
\bibfield{author}{\bibinfo{person}{Alice Kwak}, \bibinfo{person}{Cheonkam Jeong}, \bibinfo{person}{Gaetano Forte}, \bibinfo{person}{Derek Bambauer}, \bibinfo{person}{Clayton Morrison}, {and} \bibinfo{person}{Mihai Surdeanu}.} \bibinfo{year}{2023}\natexlab{}.
\newblock \showarticletitle{Information Extraction from Legal Wills: How Well Does {GPT}-4 Do?}. In \bibinfo{booktitle}{\emph{Findings of the Association for Computational Linguistics: EMNLP 2023}}, \bibfield{editor}{\bibinfo{person}{Houda Bouamor}, \bibinfo{person}{Juan Pino}, {and} \bibinfo{person}{Kalika Bali}} (Eds.). \bibinfo{publisher}{Association for Computational Linguistics}, \bibinfo{address}{Singapore}, \bibinfo{pages}{4336--4353}.
\newblock
\urldef\tempurl%
\url{https://doi.org/10.18653/v1/2023.findings-emnlp.287}
\showDOI{\tempurl}


\bibitem[Lau et~al\mbox{.}(2012)]%
        {lau2012text}
\bibfield{author}{\bibinfo{person}{Raymond~YK Lau}, \bibinfo{person}{SY Liao}, \bibinfo{person}{Ron Chi-Wai Kwok}, \bibinfo{person}{Kaiquan Xu}, \bibinfo{person}{Yunqing Xia}, {and} \bibinfo{person}{Yuefeng Li}.} \bibinfo{year}{2012}\natexlab{}.
\newblock \showarticletitle{Text mining and probabilistic language modeling for online review spam detection}.
\newblock \bibinfo{journal}{\emph{ACM Transactions on Management Information Systems (TMIS)}} \bibinfo{volume}{2}, \bibinfo{number}{4} (\bibinfo{year}{2012}), \bibinfo{pages}{1--30}.
\newblock


\bibitem[Lee et~al\mbox{.}(2019)]%
        {10.1093/bioinformatics/btz682}
\bibfield{author}{\bibinfo{person}{Jinhyuk Lee}, \bibinfo{person}{Wonjin Yoon}, \bibinfo{person}{Sungdong Kim}, \bibinfo{person}{Donghyeon Kim}, \bibinfo{person}{Sunkyu Kim}, \bibinfo{person}{Chan~Ho So}, {and} \bibinfo{person}{Jaewoo Kang}.} \bibinfo{year}{2019}\natexlab{}.
\newblock \showarticletitle{{BioBERT: a pre-trained biomedical language representation model for biomedical text mining}}.
\newblock \bibinfo{journal}{\emph{Bioinformatics}} \bibinfo{volume}{36}, \bibinfo{number}{4} (\bibinfo{date}{09} \bibinfo{year}{2019}), \bibinfo{pages}{1234--1240}.
\newblock
\showISSN{1367-4803}
\urldef\tempurl%
\url{https://doi.org/10.1093/bioinformatics/btz682}
\showDOI{\tempurl}
\showeprint{https://academic.oup.com/bioinformatics/article-pdf/36/4/1234/48983216/bioinformatics\_36\_4\_1234.pdf}


\bibitem[Lewis et~al\mbox{.}(2021)]%
        {lewis2021retrievalaugmented}
\bibfield{author}{\bibinfo{person}{Patrick Lewis}, \bibinfo{person}{Ethan Perez}, \bibinfo{person}{Aleksandra Piktus}, \bibinfo{person}{Fabio Petroni}, \bibinfo{person}{Vladimir Karpukhin}, \bibinfo{person}{Naman Goyal}, \bibinfo{person}{Heinrich Küttler}, \bibinfo{person}{Mike Lewis}, \bibinfo{person}{Wen tau Yih}, \bibinfo{person}{Tim Rocktäschel}, \bibinfo{person}{Sebastian Riedel}, {and} \bibinfo{person}{Douwe Kiela}.} \bibinfo{year}{2021}\natexlab{}.
\newblock \bibinfo{title}{Retrieval-Augmented Generation for Knowledge-Intensive NLP Tasks}.
\newblock
\newblock
\showeprint[arxiv]{2005.11401}~[cs.CL]


\bibitem[Li et~al\mbox{.}(2023)]%
        {li2023loftq}
\bibfield{author}{\bibinfo{person}{Yixiao Li}, \bibinfo{person}{Yifan Yu}, \bibinfo{person}{Chen Liang}, \bibinfo{person}{Pengcheng He}, \bibinfo{person}{Nikos Karampatziakis}, \bibinfo{person}{Weizhu Chen}, {and} \bibinfo{person}{Tuo Zhao}.} \bibinfo{year}{2023}\natexlab{}.
\newblock \bibinfo{title}{LoftQ: LoRA-Fine-Tuning-Aware Quantization for Large Language Models}.
\newblock
\newblock
\showeprint[arxiv]{2310.08659}~[cs.CL]


\bibitem[Manuel et~al\mbox{.}(2010a)]%
        {article}
\bibfield{author}{\bibinfo{person}{Douglas Manuel}, \bibinfo{person}{Laura Rosella}, {and} \bibinfo{person}{Thérèse Stukel}.} \bibinfo{year}{2010}\natexlab{a}.
\newblock \showarticletitle{Importance of accurately identifying disease in studies using electronic health records}.
\newblock \bibinfo{journal}{\emph{BMJ (Clinical research ed.)}}  \bibinfo{volume}{341} (\bibinfo{date}{08} \bibinfo{year}{2010}), \bibinfo{pages}{c4226}.
\newblock
\urldef\tempurl%
\url{https://doi.org/10.1136/bmj.c4226}
\showDOI{\tempurl}


\bibitem[Manuel et~al\mbox{.}(2010b)]%
        {manuel2010importance}
\bibfield{author}{\bibinfo{person}{Douglas~G Manuel}, \bibinfo{person}{Laura~C Rosella}, {and} \bibinfo{person}{Th{\'e}r{\`e}se~A Stukel}.} \bibinfo{year}{2010}\natexlab{b}.
\newblock \showarticletitle{Importance of accurately identifying disease in studies using electronic health records}.
\newblock \bibinfo{journal}{\emph{Bmj}}  \bibinfo{volume}{341} (\bibinfo{year}{2010}).
\newblock


\bibitem[Melucci(2009)]%
        {Melucci2009}
\bibfield{author}{\bibinfo{person}{Massimo Melucci}.} \bibinfo{year}{2009}\natexlab{}.
\newblock \bibinfo{booktitle}{\emph{Boolean Model}}.
\newblock \bibinfo{publisher}{Springer US}, \bibinfo{address}{Boston, MA}, \bibinfo{pages}{260--260}.
\newblock
\showISBNx{978-0-387-39940-9}
\urldef\tempurl%
\url{https://doi.org/10.1007/978-0-387-39940-9_917}
\showDOI{\tempurl}


\bibitem[OpenAI(2023)]%
        {openai2023gpt4}
\bibfield{author}{\bibinfo{person}{OpenAI}.} \bibinfo{year}{2023}\natexlab{}.
\newblock \bibinfo{title}{GPT-4 Technical Report}.
\newblock
\newblock
\showeprint[arxiv]{2303.08774}~[cs.CL]


\bibitem[Reimers and Gurevych(2019)]%
        {reimers-2019-sentence-bert}
\bibfield{author}{\bibinfo{person}{Nils Reimers} {and} \bibinfo{person}{Iryna Gurevych}.} \bibinfo{year}{2019}\natexlab{}.
\newblock \showarticletitle{Sentence-BERT: Sentence Embeddings using Siamese BERT-Networks}. In \bibinfo{booktitle}{\emph{Proceedings of the 2019 Conference on Empirical Methods in Natural Language Processing}}. \bibinfo{publisher}{Association for Computational Linguistics}.
\newblock
\urldef\tempurl%
\url{https://arxiv.org/abs/1908.10084}
\showURL{%
\tempurl}


\bibitem[Sammut and Webb(2010)]%
        {ref1}
\bibfield{editor}{\bibinfo{person}{Claude Sammut} {and} \bibinfo{person}{Geoffrey~I. Webb}} (Eds.). \bibinfo{year}{2010}\natexlab{}.
\newblock \bibinfo{booktitle}{\emph{TF--IDF}}.
\newblock \bibinfo{publisher}{Springer US}, \bibinfo{address}{Boston, MA}, \bibinfo{pages}{986--987}.
\newblock
\showISBNx{978-0-387-30164-8}
\urldef\tempurl%
\url{https://doi.org/10.1007/978-0-387-30164-8_832}
\showDOI{\tempurl}


\bibitem[Sedlakova et~al\mbox{.}(2023)]%
        {10.1371/journal.pdig.0000347}
\bibfield{author}{\bibinfo{person}{Jana Sedlakova}, \bibinfo{person}{Paola Daniore}, \bibinfo{person}{Andrea Horn~Wintsch}, \bibinfo{person}{Markus Wolf}, \bibinfo{person}{Mina Stanikic}, \bibinfo{person}{Christina Haag}, \bibinfo{person}{Chloé Sieber}, \bibinfo{person}{Gerold Schneider}, \bibinfo{person}{Kaspar Staub}, \bibinfo{person}{Dominik Alois~Ettlin}, \bibinfo{person}{Oliver Grübner}, \bibinfo{person}{Fabio Rinaldi}, \bibinfo{person}{Viktor von Wyl}, {and} \bibinfo{person}{for the University~of Zurich Digital Society Initiative (UZH-DSI) Health~Community}.} \bibinfo{year}{2023}\natexlab{}.
\newblock \showarticletitle{Challenges and best practices for digital unstructured data enrichment in health research: A systematic narrative review}.
\newblock \bibinfo{journal}{\emph{PLOS Digital Health}} \bibinfo{volume}{2}, \bibinfo{number}{10} (\bibinfo{date}{10} \bibinfo{year}{2023}), \bibinfo{pages}{1--22}.
\newblock
\urldef\tempurl%
\url{https://doi.org/10.1371/journal.pdig.0000347}
\showDOI{\tempurl}


\bibitem[Sedlakova et~al\mbox{.}(2022)]%
        {Sedlakova2022.07.28.22278137}
\bibfield{author}{\bibinfo{person}{Jana Sedlakova}, \bibinfo{person}{Paola Daniore}, \bibinfo{person}{Andrea~Horn Wintsch}, \bibinfo{person}{Markus Wolf}, \bibinfo{person}{Mina Stanikic}, \bibinfo{person}{Christina Haag}, \bibinfo{person}{Chlo{\'e} Sieber}, \bibinfo{person}{Gerold Schneider}, \bibinfo{person}{Kaspar Staub}, \bibinfo{person}{Dominik~Alois Ettlin}, \bibinfo{person}{Oliver Gr{\"u}bner}, \bibinfo{person}{Fabio Rinaldi}, \bibinfo{person}{Viktor von Wyl}, {and} \bibinfo{person}{University of Zurich Digital Society Initiative (UZH-DSI) Health~Community}.} \bibinfo{year}{2022}\natexlab{}.
\newblock \showarticletitle{Challenges and best practices for digital unstructured data enrichment in health research: a systematic narrative review}.
\newblock \bibinfo{journal}{\emph{medRxiv}} (\bibinfo{year}{2022}).
\newblock
\urldef\tempurl%
\url{https://doi.org/10.1101/2022.07.28.22278137}
\showDOI{\tempurl}
\showeprint{https://www.medrxiv.org/content/early/2022/07/29/2022.07.28.22278137.full.pdf}


\bibitem[Singhal et~al\mbox{.}(2023)]%
        {singhal2023expertlevel}
\bibfield{author}{\bibinfo{person}{Karan Singhal}, \bibinfo{person}{Tao Tu}, \bibinfo{person}{Juraj Gottweis}, \bibinfo{person}{Rory Sayres}, \bibinfo{person}{Ellery Wulczyn}, \bibinfo{person}{Le Hou}, \bibinfo{person}{Kevin Clark}, \bibinfo{person}{Stephen Pfohl}, \bibinfo{person}{Heather Cole-Lewis}, \bibinfo{person}{Darlene Neal}, \bibinfo{person}{Mike Schaekermann}, \bibinfo{person}{Amy Wang}, \bibinfo{person}{Mohamed Amin}, \bibinfo{person}{Sami Lachgar}, \bibinfo{person}{Philip Mansfield}, \bibinfo{person}{Sushant Prakash}, \bibinfo{person}{Bradley Green}, \bibinfo{person}{Ewa Dominowska}, \bibinfo{person}{Blaise~Aguera y Arcas}, \bibinfo{person}{Nenad Tomasev}, \bibinfo{person}{Yun Liu}, \bibinfo{person}{Renee Wong}, \bibinfo{person}{Christopher Semturs}, \bibinfo{person}{S.~Sara Mahdavi}, \bibinfo{person}{Joelle Barral}, \bibinfo{person}{Dale Webster}, \bibinfo{person}{Greg~S. Corrado}, \bibinfo{person}{Yossi Matias}, \bibinfo{person}{Shekoofeh Azizi}, \bibinfo{person}{Alan Karthikesalingam}, {and}
  \bibinfo{person}{Vivek Natarajan}.} \bibinfo{year}{2023}\natexlab{}.
\newblock \bibinfo{title}{Towards Expert-Level Medical Question Answering with Large Language Models}.
\newblock
\newblock
\showeprint[arxiv]{2305.09617}~[cs.CL]


\bibitem[Sontag et~al\mbox{.}(2012)]%
        {sontag2012probabilistic}
\bibfield{author}{\bibinfo{person}{David Sontag}, \bibinfo{person}{Kevyn Collins-Thompson}, \bibinfo{person}{Paul~N Bennett}, \bibinfo{person}{Ryen~W White}, \bibinfo{person}{Susan Dumais}, {and} \bibinfo{person}{Bodo Billerbeck}.} \bibinfo{year}{2012}\natexlab{}.
\newblock \showarticletitle{Probabilistic models for personalizing web search}. In \bibinfo{booktitle}{\emph{Proceedings of the fifth ACM international conference on Web search and data mining}}. \bibinfo{pages}{433--442}.
\newblock


\bibitem[Taori et~al\mbox{.}(2023)]%
        {alpaca}
\bibfield{author}{\bibinfo{person}{Rohan Taori}, \bibinfo{person}{Ishaan Gulrajani}, \bibinfo{person}{Tianyi Zhang}, \bibinfo{person}{Yann Dubois}, \bibinfo{person}{Xuechen Li}, \bibinfo{person}{Carlos Guestrin}, \bibinfo{person}{Percy Liang}, {and} \bibinfo{person}{Tatsunori~B. Hashimoto}.} \bibinfo{year}{2023}\natexlab{}.
\newblock \bibinfo{title}{Stanford Alpaca: An Instruction-following LLaMA model}.
\newblock \bibinfo{howpublished}{\url{https://github.com/tatsu-lab/stanford_alpaca}}.
\newblock


\bibitem[Torregrossa et~al\mbox{.}(2021)]%
        {torregrossa2021survey}
\bibfield{author}{\bibinfo{person}{Fran{\c{c}}ois Torregrossa}, \bibinfo{person}{Robin Allesiardo}, \bibinfo{person}{Vincent Claveau}, \bibinfo{person}{Nihel Kooli}, {and} \bibinfo{person}{Guillaume Gravier}.} \bibinfo{year}{2021}\natexlab{}.
\newblock \showarticletitle{A survey on training and evaluation of word embeddings}.
\newblock \bibinfo{journal}{\emph{International Journal of Data Science and Analytics}}  \bibinfo{volume}{11} (\bibinfo{year}{2021}), \bibinfo{pages}{85--103}.
\newblock


\bibitem[Touvron et~al\mbox{.}(2023)]%
        {touvron2023llama}
\bibfield{author}{\bibinfo{person}{Hugo Touvron}, \bibinfo{person}{Thibaut Lavril}, \bibinfo{person}{Gautier Izacard}, \bibinfo{person}{Xavier Martinet}, \bibinfo{person}{Marie-Anne Lachaux}, \bibinfo{person}{Timothée Lacroix}, \bibinfo{person}{Baptiste Rozière}, \bibinfo{person}{Naman Goyal}, \bibinfo{person}{Eric Hambro}, \bibinfo{person}{Faisal Azhar}, \bibinfo{person}{Aurelien Rodriguez}, \bibinfo{person}{Armand Joulin}, \bibinfo{person}{Edouard Grave}, {and} \bibinfo{person}{Guillaume Lample}.} \bibinfo{year}{2023}\natexlab{}.
\newblock \bibinfo{title}{LLaMA: Open and Efficient Foundation Language Models}.
\newblock
\newblock
\showeprint[arxiv]{2302.13971}~[cs.CL]


\bibitem[Vaswani et~al\mbox{.}(2017)]%
        {NIPS2017_3f5ee243}
\bibfield{author}{\bibinfo{person}{Ashish Vaswani}, \bibinfo{person}{Noam Shazeer}, \bibinfo{person}{Niki Parmar}, \bibinfo{person}{Jakob Uszkoreit}, \bibinfo{person}{Llion Jones}, \bibinfo{person}{Aidan~N Gomez}, \bibinfo{person}{\L~ukasz Kaiser}, {and} \bibinfo{person}{Illia Polosukhin}.} \bibinfo{year}{2017}\natexlab{}.
\newblock \showarticletitle{Attention is All you Need}. In \bibinfo{booktitle}{\emph{Advances in Neural Information Processing Systems}}, \bibfield{editor}{\bibinfo{person}{I.~Guyon}, \bibinfo{person}{U.~Von Luxburg}, \bibinfo{person}{S.~Bengio}, \bibinfo{person}{H.~Wallach}, \bibinfo{person}{R.~Fergus}, \bibinfo{person}{S.~Vishwanathan}, {and} \bibinfo{person}{R.~Garnett}} (Eds.), Vol.~\bibinfo{volume}{30}. \bibinfo{publisher}{Curran Associates, Inc.}
\newblock
\urldef\tempurl%
\url{https://proceedings.neurips.cc/paper_files/paper/2017/file/3f5ee243547dee91fbd053c1c4a845aa-Paper.pdf}
\showURL{%
\tempurl}


\bibitem[Wu et~al\mbox{.}(2024)]%
        {doi:10.1056/AIdbp2300092}
\bibfield{author}{\bibinfo{person}{Sean Wu}, \bibinfo{person}{Michael Koo}, \bibinfo{person}{Lesley Blum}, \bibinfo{person}{Andy Black}, \bibinfo{person}{Liyo Kao}, \bibinfo{person}{Zhe Fei}, \bibinfo{person}{Fabien Scalzo}, {and} \bibinfo{person}{Ira Kurtz}.} \bibinfo{year}{2024}\natexlab{}.
\newblock \showarticletitle{Benchmarking Open-Source Large Language Models, GPT-4 and Claude 2 on Multiple-Choice Questions in Nephrology}.
\newblock \bibinfo{journal}{\emph{NEJM AI}} \bibinfo{volume}{1}, \bibinfo{number}{2} (\bibinfo{year}{2024}), \bibinfo{pages}{AIdbp2300092}.
\newblock
\urldef\tempurl%
\url{https://doi.org/10.1056/AIdbp2300092}
\showDOI{\tempurl}
\showeprint{https://ai.nejm.org/doi/pdf/10.1056/AIdbp2300092}


\bibitem[Xiao et~al\mbox{.}(2023)]%
        {pmlr-v202-xiao23c}
\bibfield{author}{\bibinfo{person}{Guangxuan Xiao}, \bibinfo{person}{Ji Lin}, \bibinfo{person}{Mickael Seznec}, \bibinfo{person}{Hao Wu}, \bibinfo{person}{Julien Demouth}, {and} \bibinfo{person}{Song Han}.} \bibinfo{year}{2023}\natexlab{}.
\newblock \showarticletitle{{S}mooth{Q}uant: Accurate and Efficient Post-Training Quantization for Large Language Models}. In \bibinfo{booktitle}{\emph{Proceedings of the 40th International Conference on Machine Learning}} \emph{(\bibinfo{series}{Proceedings of Machine Learning Research}, Vol.~\bibinfo{volume}{202})}, \bibfield{editor}{\bibinfo{person}{Andreas Krause}, \bibinfo{person}{Emma Brunskill}, \bibinfo{person}{Kyunghyun Cho}, \bibinfo{person}{Barbara Engelhardt}, \bibinfo{person}{Sivan Sabato}, {and} \bibinfo{person}{Jonathan Scarlett}} (Eds.). \bibinfo{publisher}{PMLR}, \bibinfo{pages}{38087--38099}.
\newblock
\urldef\tempurl%
\url{https://proceedings.mlr.press/v202/xiao23c.html}
\showURL{%
\tempurl}


\bibitem[Xu et~al\mbox{.}(2023)]%
        {xu2023wizardlm}
\bibfield{author}{\bibinfo{person}{Can Xu}, \bibinfo{person}{Qingfeng Sun}, \bibinfo{person}{Kai Zheng}, \bibinfo{person}{Xiubo Geng}, \bibinfo{person}{Pu Zhao}, \bibinfo{person}{Jiazhan Feng}, \bibinfo{person}{Chongyang Tao}, {and} \bibinfo{person}{Daxin Jiang}.} \bibinfo{year}{2023}\natexlab{}.
\newblock \bibinfo{title}{WizardLM: Empowering Large Language Models to Follow Complex Instructions}.
\newblock
\newblock
\showeprint[arxiv]{2304.12244}~[cs.CL]


\bibitem[Xu et~al\mbox{.}(2020)]%
        {XU20201}
\bibfield{author}{\bibinfo{person}{Dongliang Xu}, \bibinfo{person}{Zhihong Tian}, \bibinfo{person}{Rufeng Lai}, \bibinfo{person}{Xiangtao Kong}, \bibinfo{person}{Zhiyuan Tan}, {and} \bibinfo{person}{Wei Shi}.} \bibinfo{year}{2020}\natexlab{}.
\newblock \showarticletitle{Deep learning based emotion analysis of microblog texts}.
\newblock \bibinfo{journal}{\emph{Information Fusion}}  \bibinfo{volume}{64} (\bibinfo{year}{2020}), \bibinfo{pages}{1--11}.
\newblock
\showISSN{1566-2535}
\urldef\tempurl%
\url{https://doi.org/10.1016/j.inffus.2020.06.002}
\showDOI{\tempurl}


\bibitem[Yao et~al\mbox{.}(2023)]%
        {yao2023zeroquantv2}
\bibfield{author}{\bibinfo{person}{Zhewei Yao}, \bibinfo{person}{Xiaoxia Wu}, \bibinfo{person}{Cheng Li}, \bibinfo{person}{Stephen Youn}, {and} \bibinfo{person}{Yuxiong He}.} \bibinfo{year}{2023}\natexlab{}.
\newblock \bibinfo{title}{ZeroQuant-V2: Exploring Post-training Quantization in LLMs from Comprehensive Study to Low Rank Compensation}.
\newblock
\newblock
\showeprint[arxiv]{2303.08302}~[cs.LG]


\bibitem[Yue et~al\mbox{.}(2021a)]%
        {yue2021annotated}
\bibfield{author}{\bibinfo{person}{Xiang Yue}, \bibinfo{person}{Xinliang~Frederick Zhang}, {and} \bibinfo{person}{Huan Sun}.} \bibinfo{year}{2021}\natexlab{a}.
\newblock \bibinfo{title}{Annotated Question-Answer Pairs for Clinical Notes in the MIMIC-III Database}.
\newblock \bibinfo{howpublished}{(version 1.0.0)}.
\newblock
\urldef\tempurl%
\url{https://doi.org/10.13026/j0y6-bw05}
\showURL{%
\tempurl}


\bibitem[Yue et~al\mbox{.}(2021b)]%
        {yue2021cliniqg4qa}
\bibfield{author}{\bibinfo{person}{Xiang Yue}, \bibinfo{person}{Xinliang~Frederick Zhang}, \bibinfo{person}{Ziyu Yao}, \bibinfo{person}{Simon Lin}, {and} \bibinfo{person}{Huan Sun}.} \bibinfo{year}{2021}\natexlab{b}.
\newblock \bibinfo{title}{CliniQG4QA: Generating Diverse Questions for Domain Adaptation of Clinical Question Answering}.
\newblock
\newblock
\showeprint[arxiv]{2010.16021}~[cs.CL]


\bibitem[Zhao et~al\mbox{.}(2023)]%
        {zhao2023survey}
\bibfield{author}{\bibinfo{person}{Wayne~Xin Zhao}, \bibinfo{person}{Kun Zhou}, \bibinfo{person}{Junyi Li}, \bibinfo{person}{Tianyi Tang}, \bibinfo{person}{Xiaolei Wang}, \bibinfo{person}{Yupeng Hou}, \bibinfo{person}{Yingqian Min}, \bibinfo{person}{Beichen Zhang}, \bibinfo{person}{Junjie Zhang}, \bibinfo{person}{Zican Dong}, \bibinfo{person}{Yifan Du}, \bibinfo{person}{Chen Yang}, \bibinfo{person}{Yushuo Chen}, \bibinfo{person}{Zhipeng Chen}, \bibinfo{person}{Jinhao Jiang}, \bibinfo{person}{Ruiyang Ren}, \bibinfo{person}{Yifan Li}, \bibinfo{person}{Xinyu Tang}, \bibinfo{person}{Zikang Liu}, \bibinfo{person}{Peiyu Liu}, \bibinfo{person}{Jian-Yun Nie}, {and} \bibinfo{person}{Ji-Rong Wen}.} \bibinfo{year}{2023}\natexlab{}.
\newblock \bibinfo{title}{A Survey of Large Language Models}.
\newblock
\newblock
\showeprint[arxiv]{2303.18223}~[cs.CL]


\bibitem[Zheng et~al\mbox{.}(2023)]%
        {zheng2023judging}
\bibfield{author}{\bibinfo{person}{Lianmin Zheng}, \bibinfo{person}{Wei-Lin Chiang}, \bibinfo{person}{Ying Sheng}, \bibinfo{person}{Siyuan Zhuang}, \bibinfo{person}{Zhanghao Wu}, \bibinfo{person}{Yonghao Zhuang}, \bibinfo{person}{Zi Lin}, \bibinfo{person}{Zhuohan Li}, \bibinfo{person}{Dacheng Li}, \bibinfo{person}{Eric.~P Xing}, \bibinfo{person}{Hao Zhang}, \bibinfo{person}{Joseph~E. Gonzalez}, {and} \bibinfo{person}{Ion Stoica}.} \bibinfo{year}{2023}\natexlab{}.
\newblock \bibinfo{title}{Judging LLM-as-a-judge with MT-Bench and Chatbot Arena}.
\newblock
\newblock
\showeprint[arxiv]{2306.05685}~[cs.CL]


\bibitem[Zhu et~al\mbox{.}(2023)]%
        {zhu2023large}
\bibfield{author}{\bibinfo{person}{Yutao Zhu}, \bibinfo{person}{Huaying Yuan}, \bibinfo{person}{Shuting Wang}, \bibinfo{person}{Jiongnan Liu}, \bibinfo{person}{Wenhan Liu}, \bibinfo{person}{Chenlong Deng}, \bibinfo{person}{Zhicheng Dou}, {and} \bibinfo{person}{Ji-Rong Wen}.} \bibinfo{year}{2023}\natexlab{}.
\newblock \bibinfo{title}{Large Language Models for Information Retrieval: A Survey}.
\newblock
\newblock
\showeprint[arxiv]{2308.07107}~[cs.CL]


\end{thebibliography}


\end{document}